\documentclass[a4paper,11pt]{article}
\usepackage{pos}
\usepackage{subcaption}
\usepackage{graphicx}

\title{Bayesian Inference for Contemporary Lattice Quantum Field Theory}
\ShortTitle{Bayesian Inference for Contemporary Lattice QFT}

\author*[a]{Julien Frison}

\affiliation{John von Neumann-Institut f\"ur Computing NIC, Deutsches Elektronen-Synchrotron DESY, Platanenallee 6, 15738 Zeuthen, Germany}
\emailAdd{julien.frison@desy.de}

\abstract{Bayesian inference provides a rigorous framework to encapsulate our knowledge and uncertainty
regarding various physical quantities in a well-defined and self-contained manner. Utilising modern
tools, such Bayesian models can be constructed with a remarkable flexibility, leaving us totally free
to carefully choose which assumption should be strictly enforced and which should on the contrary
be relaxed. The practical evaluation of these assumptions, together with the data-driven selection
or averaging of models, also appears in a very natural way.
In this presentation, I discuss its application in the context of lattice QCD and its common
statistical problems. As a concrete illustration, I present a few parametric and non-parametric
hierarchical models applied to actual correlator data, from single exponential fits to spectral functions.}

\FullConference{The 40th International Symposium on Lattice Field Theory (Lattice 2023)\\
July 31st - August 4th, 2023\\
Fermi National Accelerator Laboratory\\}

\begin{document}
\maketitle

\section{Introduction}

Probabilistic programming languages (PPL) have made huge progress in the last decade, in large part thanks to new Monte-Carlo samplers \cite{Hoffman2014NUTS} derived from those originally developped
for Lattice QCD \cite{Duane1987}. Combined with modern computing resources and Machine Learning libraries \cite{thetheanodevelopmentteam2016theano}, it opens the door to the automatic evaluation
of Bayesian Networks of arbitrary complexity. PPL usually implement various methods, and in particular it also connects to other Machine Learning techniques such as VAE \cite{VAE} and normalising flows \cite{NFlow}, which are often justified theoretically as a good approximation of a Bayesian inference.

To this day, this has been widely applied to epidemiology \cite{Imperial2020}, finance, geology, pharmaceutical industry, marketing, social sciences and many other fields. However, applications to LQCD are scarce, and to our knowledge mostly limited to \cite{bayesPaper,Rothkopf}.

Instead, LQCD is mostly based on a combination of traditional statistical methods developped in the 70's or earlier. This includes using the Central Limit Theorem on Markov Chains \cite{Bernstein27}, Generalised Least Square \cite{Aitken36}, Delta method \cite{Doob35}, Jacknife \cite{Quenouille56} and Bootstrapping \cite{EfronBootstrap}.
The main exception in this picture is the successful adoption of (pseudo-)Bayesian Model Averaging \cite{JayNeil} by a large share of the community.

The objective of this work is to learn physical parameters or functions, in a numerically efficient way, which extracts as much information as possible from our data, built on a theoretically
well-defined probabilistic interpretation.
This should additionally provide a unified and consistent framework describing the whole process from the data to the final result, combining as much as possible the strengths of all the different methods
currently used in the LQCD state-of-the-art. As a result, one can get rid of unnecessary assumptions and gain flexibility on model building.
Eventually, an important feature of this Bayesian point-of-view is the ability to provide interpretable metrics to test afterwards which assumptions should be used or not.

In Sec.~\ref{sec:sota} we will summarise the state-of-the-art and discuss how one can expect this to improve from Bayesian inference.
Sec.~\ref{sec:bayes} will then introduce the formalism and tools at our disposal.
Eventually, the main discussion in Sec.~\ref{sec:models} will focus on giving an overview of some models illustrating the diversity of problems which can be unified into a single framework.

\section{Current methods and beyond}
\subsection{Current methods}
\label{sec:sota}

State-of-the-art LQCD analyses are mostly organised in three somewhat independent layers, which can sometimes conflict with each other.

Its foundation usually consist in a resampling, either block\cite{10.1214/aos/1176347265} bootstrapping or jacknife. Some practioners prefer using the $\Gamma$ method \cite{GammaMethod}, and
the best choice depends on the problem considered. 
While bootstrapping is an exact non-parametric method for non-linear functions of non-Gaussian data, the $\Gamma$ method has a demonstrably faster convergence for
strongly auto-correlated data.

The next layer consists in a $\chi^2$ fit, which is essentially a maximum likelihood estimator with a Gaussian likelihood.
One practical difficulty is the need for the covariance to be known with sufficiently good precision {\it before} fitting, and the lack of a clear definition for what
{\it sufficiently good} means. 
Another difficulty is that (non-regularised) non-linear fits are often unstable for a moderately large number of parameters. Convergence is usually not theoretically guaranteed, and it is a fortiori not
guaranteed to converge to something meaningful given only a finite amount of data.

The last piece is the Akaike Information Criterion\cite{Akaike}, to give a more rigorous meaning to the results of the previous layers.
However, this requires prior knowledge of the correlated $\chi^2$, which is often not reliable.
Additionally, in principle it requires the guarantee that one model parametrises the truth.
Finally, evaluating many models over many boostrap samples tends to be a computationally intensive task, even when only a few models contribute significantly to the results.

\subsection{A fully Bayesian solution}

In this work we replace all of this with a single method which is fully Bayesian from the start to the end. 
It natively provides distributions and confidence intervals for every single object, with an automatic propagation of all errors, instead of wrapping a posteriori a point estimate inside an extra layer.
Every assumption is encoded into the choice of model, while the same framework is used for arbitrarily complicated models or choices of models covering different regimes.
These models can be evaluated with an HMC, which might take time to converge in practice but is theoretically guaranteed to converge into a well-defined posterior distribution.
Eventually, widely applicable criteria\cite{Watanabe,LOO} exist to evaluate the distance from the model to the truth, making model averaging and selection more reliable.

\section{Bayesian formalism}
\label{sec:bayes}
\subsection{The Bayes formula}
\label{sec:bayesformula}
Our inference is based on an ``inversion'' of probabilities with the Bayes formula
\begin{equation}
  P\left(a\mid y,M\right) = \frac{P\left(y\mid a,M\right)P\left(a\mid M\right)}{P\left(y\mid M\right)}.
\end{equation}
In the usual terminology, the left hand side is called the posterior distribution and expresses the probability distribution of a set of parameters $a$ in the model $M$ when a data sample $y$ is known.
This distribution is our final result, from which one can extract quantiles and other estimators to form our intervals of confidence. In the right hand side, the numerator is made of a likelihood multiplied
by a prior, both being chosen by the modeller. The denominator, called marginal likelihood or evidence, is usually ignored.

Starting from these elements, one can also define the Posterior Predictive Distribution, which expresses a Bayesian model as a generative machine-learning model:
\begin{equation}
  P\left(y'\mid y,M\right) = \int P\left(y'\mid a,M\right)P\left(a\mid y,M\right) \mathrm{d}a.
\end{equation}

\subsection{Applying the HMC}

A close formula is usually not needed for the posterior, and PPL are specifically made to deal with situations where such a manual derivation is not possible.
Configurations of the parameters $a$ can be sampled from an HMC-based algorithm, similarly to how lattice configurations are drawn.
Pushing the analogy, the negative log-likelihood (for instance a $\chi^2$ in a fit) corresponds to the choice of a lattice action, while the marginal distribution corresponds to the partition function.
A major difference is that here we do not need to manually provide an expression for the HMC forces.

\subsection{Software}

Many languages and libraries exist for Probabilistic Programming. We chose to base this work on PyMC\cite{PyMC}, with its vectorisation and automatic differentiation included through PyTensor.
With this tools at our disposal, the task for the user is usually limited to writing models in the form of a mathematical expression. 
These are not restricted to comply with any specific approximation, but only to provide a tree which compiles. 

Some analytical computations from the user can be useful, for marginalisation in particular, but only if a speedup is needed. 
Models can either be parametric (like traditional $\chi^2$ fits) or non-parametric (Bayesian Bootstrap, Gaussian Processes, Bayesian Neural Networks, \dots), and parametric models do not need to have
a positive number of degrees of freedom.
Finally, it always provides an estimate of the Kullback-Leibler divergence, a goodness-of-fit indicator which (unlike frequentist arguments or the Akaike IC) is not conditioned on the model being {\it true}. So in principle the distance from the model to the truth can be checked from a strongly data-driven point-of-view.

We will show some graphical representations of Bayesian Networks in the commonly used
{\it plate notation}, based on the Graphviz library.


\section{A few models}
\label{sec:models}
\subsection{Data}
All the models presented here are tested on some arbitrary correlator data for illustration. Those come from the CLS ensemble H101, with heavy valence quarks.
This data is provided on \cite{zenodo} alongside with the code to reproduce our results.
This is not the only type of data on which these methods would apply, and we could perfectly have decided to perform for instance chiral and continuum fits, or combine lattice and experimental data.
Instead, we chose to show the diversity of models that one can build even from the very same data.

\subsection{Bayesian Bootstrap}

\begin{figure}
  \centering
  \begin{subfigure}[b]{0.35\textwidth}
    \centering
    \includegraphics[width=0.7\textwidth]{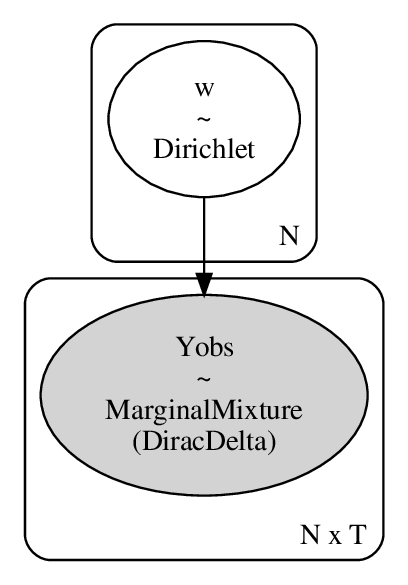}
    \caption{Bayesian Bootstrap in {\it plate notation}}
    \label{fig:graph-BB}
  \end{subfigure}
  \hfill
  \begin{subfigure}[b]{0.6\textwidth}
    \centering
    \includegraphics[width=0.8\linewidth]{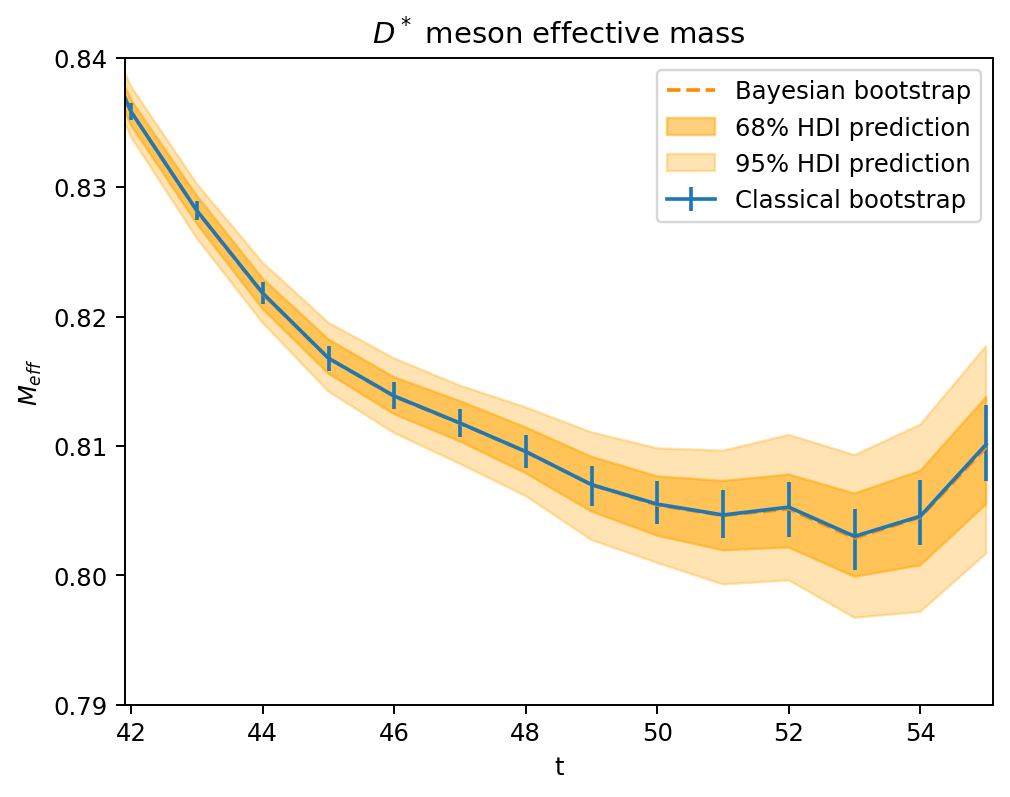}
    \caption{Effective mass of a meson}
    \label{fig:meff-BB}
  \end{subfigure}
  \caption{Left: Bayesian Bootstrap model for the observed data $Y_{obs}$ made of $N$ gauge configurations and $T$ time slices. The parameters $w$ to be inferred are associated with a Dirichlet prior so that they act as probabilities summing to one.
  Right: We show the estimate of the classical bootstrap together with the Posterior Predictive Distribution of the Bayesian bootstrap. The two methods are very similar but the Bayesian bootstrap is expected to have a better convergence.}
\end{figure}

As this is usually the first step in a LQCD analysis, reproducing the bootstrap is interesting. It has been noticed very early\cite{Rubin81} that the classical boostrap is closely related to
a Bayesian model. This is actually one of the simplest example of non-parametric model, whose graph is show in Fig.~\ref{fig:graph-BB}. It is an extreme came of Dirichlet Process Mixture, where the base
functions are Dirac deltas.
In Fig.~\ref{fig:meff-BB} we show a comparison with a classical bootstrap.

\subsection{Infering a covariance}

Performing a Generalised Least Square is a notoriously delicate task in LQCD. Indeed, the {\it true} covariance is not directly accessible and the {\it sample} covariance is a noisy estimate
which can be ill-conditioned. In the limit when the fit dimension becomes larger than the number of independent configurations, the sample covariance is even {\it guaranteed} to be singular and the $\chi^2$ can never be built. 
However, this error is usually neglected in state-of-the-art analyses.

Regularisation (eg. Ledoit-Wolf) is a partial solution to this issue, but it does not propagate uncertainties to the final result. Less importantly, it does not allow a specific choice of model to guide the covariance posterior.

\begin{figure}
  \centering
  \begin{subfigure}[b]{0.34\textwidth}
    \centering
    \includegraphics[width=0.9\textwidth]{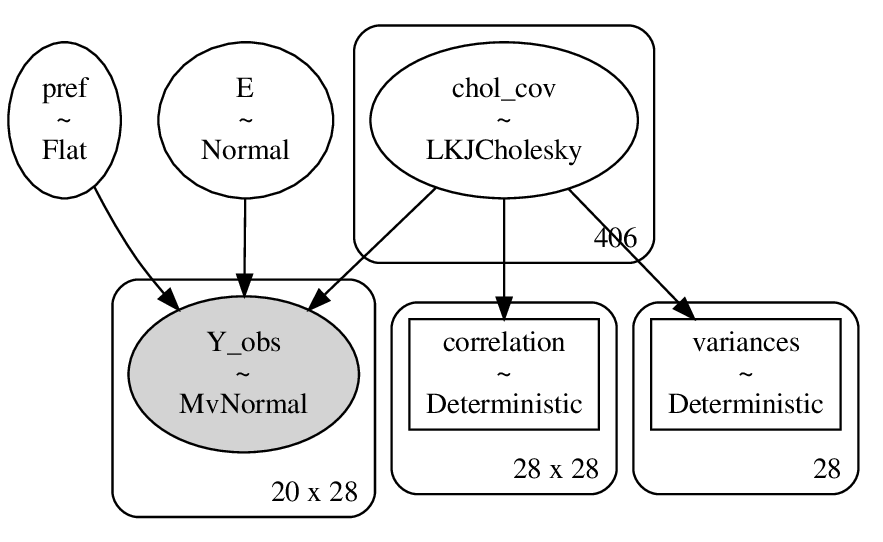}
    \caption{Exponential model with unknown covariance. An uninformative LKJ prior is used instead of a Wishart prior for practical reasons.
    One should in principle check that the result here is independent of the choice of prior.}
    \label{fig:graph-covariance}
  \end{subfigure}
  \hfill
  \begin{subfigure}[b]{0.65\textwidth}
    \centering
    \includegraphics[width=1.0\textwidth]{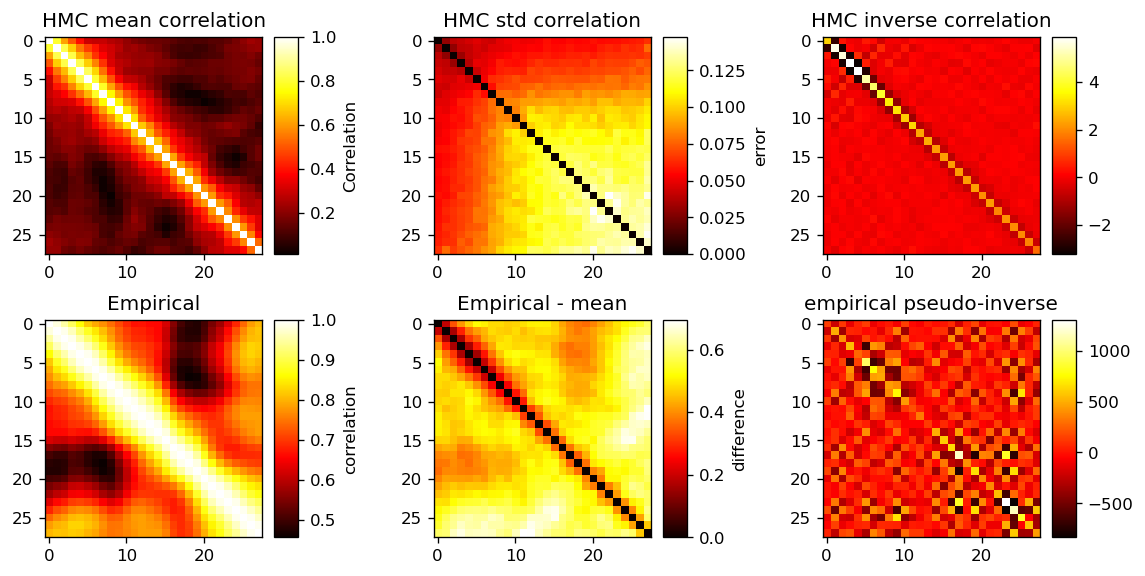}
    \caption{HMC result for the correlation and comparison with empirical correlation}
    \label{fig:heat-covariance}
  \end{subfigure}
  \caption{Left: Representation of a single-exponential model with unknown covariance for $N=20$ bins and $T=28$ time slices. 
  Top-right: we show the mean and standard deviation of the posterior distribution of the correlation matrix, then its inverse. Bottom-right: As a comparison we show the empirical correlation we were starting with, its deviation from the mean of our HMC results, and its ill-conditioned pseudo-inverse.}
\end{figure}

On the other hand, this is handled very naturally in a Bayesian framework, by promoting the covariance into a model parameter as represented in Fig.~\ref{fig:graph-covariance}.
Non-positive matrices are simply not allowed by the prior. Near-singular matrices are suppressed by the large $\chi^2$. And finally a Wishart prior (the conjugate for a Gaussian likelihood) provides shrinkage
around its scale matrix. Results are show in Fig.~\ref{fig:heat-covariance} with non-informative priors.

\subsection{Multi-exponential model with correlations}
\label{sec:multiexp}

\begin{figure}
  \centering
  \begin{subfigure}[b]{0.39\textwidth}
    \centering
    \includegraphics[width=0.9\textwidth]{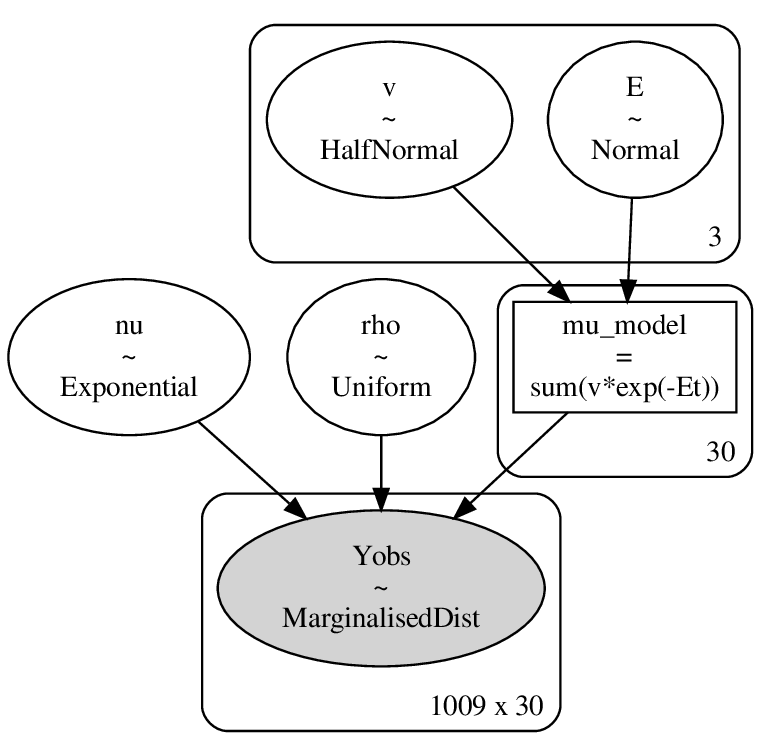}
    \caption{3-exponential model with both autocorrelation (through the lag parameters $\rho$) and time correlation (through the hyper-parameter $\nu$ of the marginalised Wishart prior)}
    \label{fig:graph-AR}
  \end{subfigure}
  \hfill
  \begin{subfigure}[b]{0.6\textwidth}
    \centering
    \includegraphics[width=1.0\textwidth]{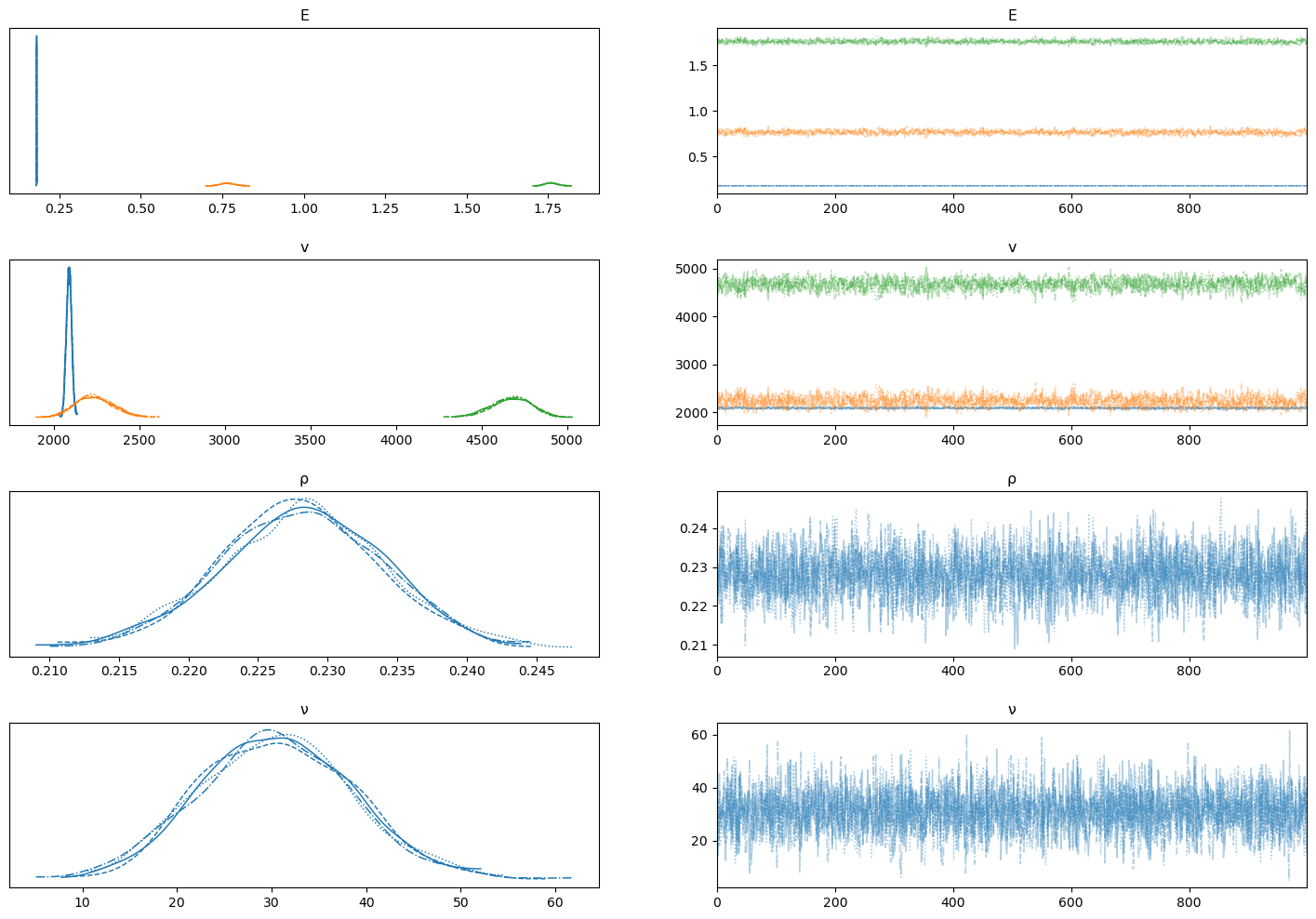}
    \caption{HMC trace, where the value of each parameter is shown as a function of the configuration number}
    \label{fig:trace-AR}
  \end{subfigure}
  \caption{Left: Model used where $Y_{obs}$ is the observed correlator on $T=30$ time slices for $N=1009$ configurations. 
  Middle: Posterior ditribution of each free parameter (one colour per component when it is a vector). Good signal is obtained on each parameter, and the three states are well-separated.
  Right: MCMC trace corresponding to the posterior, where the x-axis runs over Monte-Carlo samples and the y-axis represents the value taken by the parameter. }
\end{figure}

Multi-exponential models with time correlations plus auto-correlations had already been presented in \cite{bayesPaper} where we discuss a marginalisation strategy to facilitate the sampling.
Auto-correlations are implemented through an auto-regressive process, bearing some similarities with the $\Gamma$ method. 
Choosing as a first step a simplified VAR model in which the auto-correlation modes are independent of time, the auto-regression reads
\begin{equation}
y_{\tau}(t) = \rho_0 + \sum_{i=1}^r \rho_i y_{\tau-i}(t) + \xi_{\tau}(t),\qquad \langle\xi_{\tau}(t)\xi_{\tau}(t')\rangle\not=0.
\end{equation}
An example is shown in Fig.~\ref{fig:graph-AR} and its results in Fig.~\ref{fig:trace-AR}.

\subsection{Model averaging}

\begin{figure}
  \centering
  \includegraphics[width=0.5\textwidth]{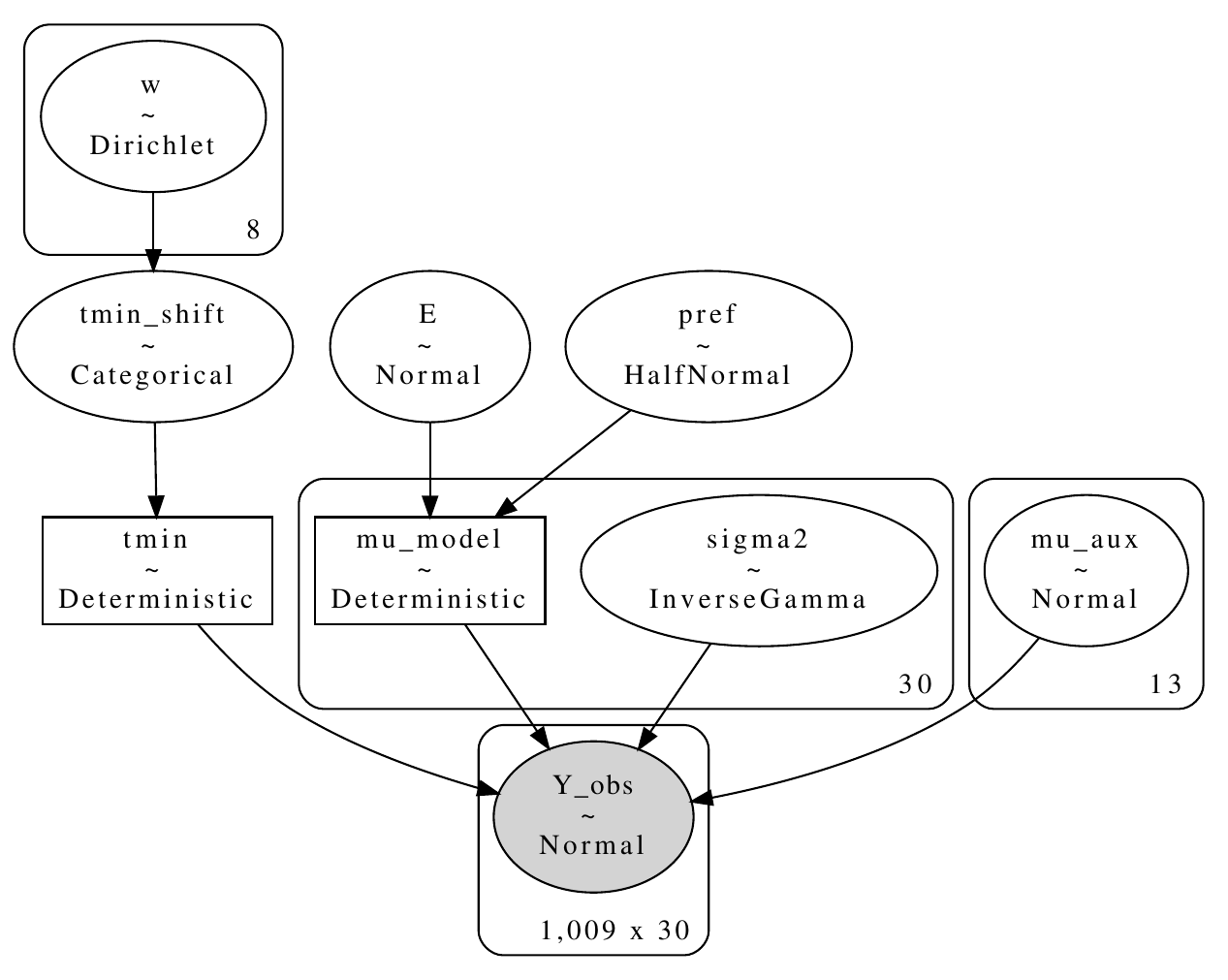}
  \caption{Model used for the Bayesian Model Average, when $t_{min}$ is not marginalised. The auxiliary parameters $\mu_{aux}$ describe a cut, while multi-exponential $\mu_{model}$ is only used for times in the plateau determined by $t_{min}$. This allows to fix the time extent of the observed data $Y_{obs}$ and therefore make meaningful comparisons where the posterior distribution for each sub-model (fixed $t_{min}$) is conditioned with the same data.}
  \label{fig:graph-BMA}
\end{figure}

\begin{figure}
  \centering
  \begin{subfigure}[b]{0.55\textwidth}
    \centering
    \includegraphics[width=0.99\textwidth]{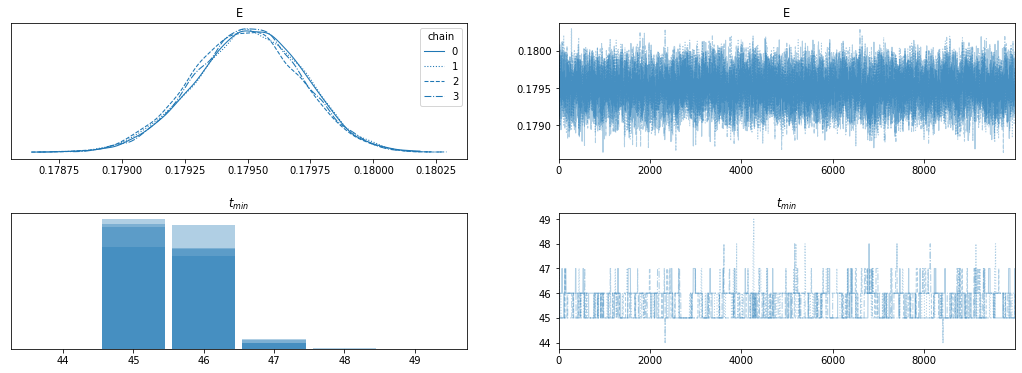}
    \caption{Posterior distribution and trace of $(E,t_{min})$. As we chose the non-marginalised model, it is important to check that $t_{min}$ moves sufficiently.}
  \end{subfigure}
  \begin{subfigure}[b]{0.4\textwidth}
    \centering
    \includegraphics[width=0.95\textwidth]{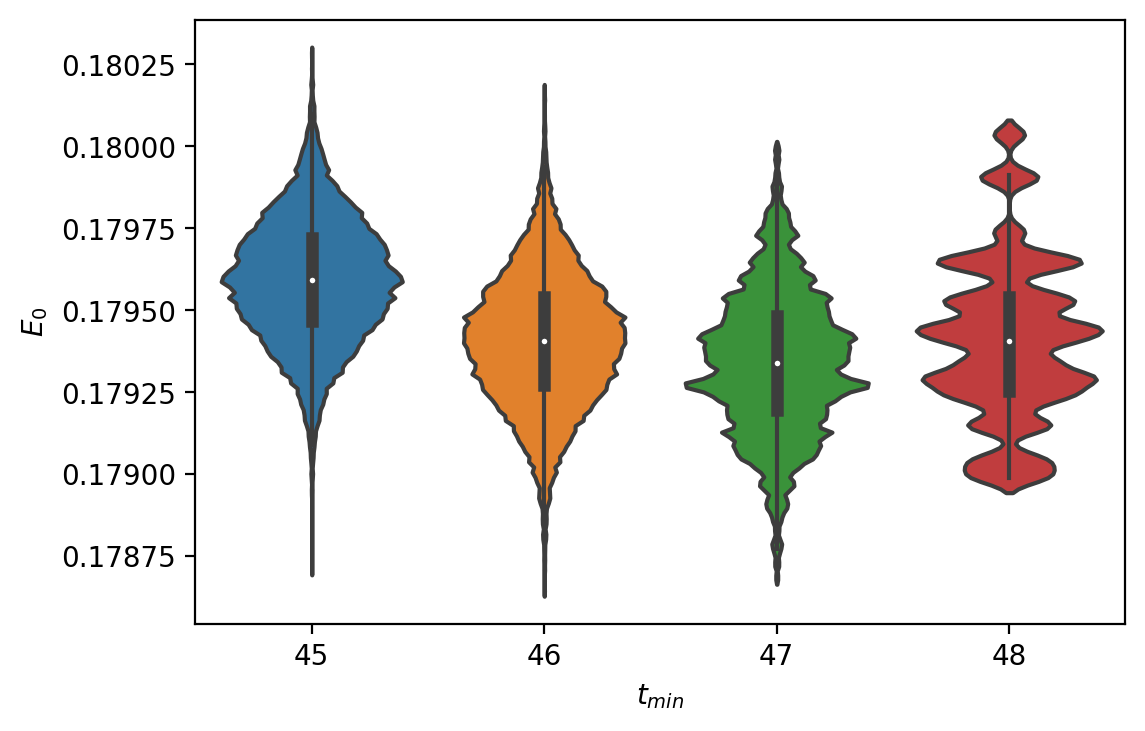}
    \caption{Posterior distribution of $E$ conditioned with a given $t_{min}$, as a violin plot}
  \end{subfigure}
  \caption{The bell-shaped posterior distribution of $E$ on the left is directly obtained as a combination of all values of $t_{min}$, together with the posterior probability of each $t_{min}$.
  One can however look at subsets of the trace to construct conditional distributions as shown on the right, and check that all subsets with high $t_{min}$ probability give compatible results.
  This also makes it obvious that this method spends more time sampling the most relevant $t_{min}$ sectors so that the result is smoother there.}
  \label{fig:trace-BMA}
\end{figure}

A fully Bayesian model provides many ways to average models, including some presented in \cite{bayesPaper}. A particularly interesting one is the construction of a supermodel which contains all the base models.
This can be obtained by adding a discrete latent variable as in the model of Fig.~\ref{fig:graph-BMA}, whose results are shown in Fig.~\ref{fig:trace-BMA}. This variable can trivially be marginalised if preferred.

This is a genuine BMA, based on marginal distributions, and avoids many discussions on the choice of an Information Criterion and its asymptotic validity. It also allows to sample all models at the same time
and spend more time on sampling the most relevant models, which means the computing cost scales better than Akaike-based methods for large model sets. 

\subsection{Spectral function}

\begin{figure}
  \centering
  \begin{subfigure}[b]{0.39\textwidth}
    \centering
    \includegraphics[width=0.7\textwidth]{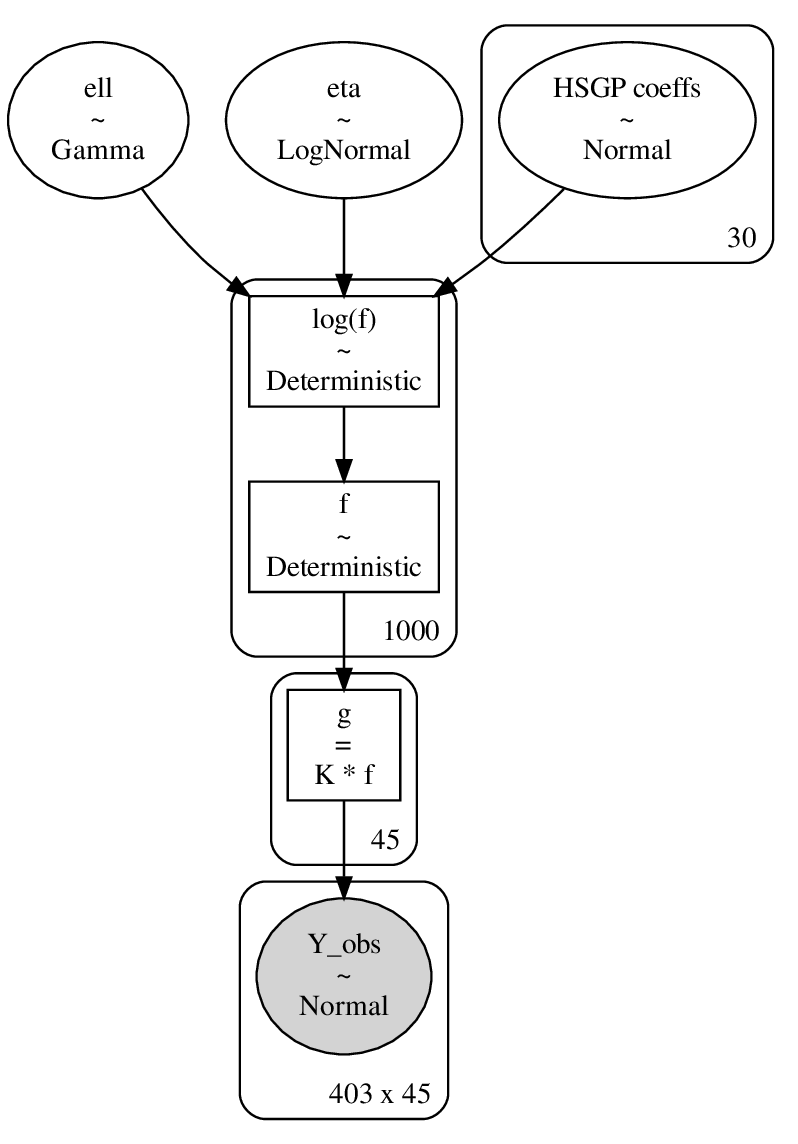}
    \caption{$\eta$ and $\ell$ are the parameters of the Mat\'ern kernel for the GP prior, which are approximated with a HSGP \cite{HSGP}.
    The correlator is reconstructed with Simpson integration using the Laplace kernel $K$ and the actual correlator is observed on $N=403$ configurations and $T=45$ time slices.}
  \end{subfigure}
  \hfill
  \begin{subfigure}[b]{0.60\textwidth}
    \centering
    \includegraphics[width=1.0\textwidth]{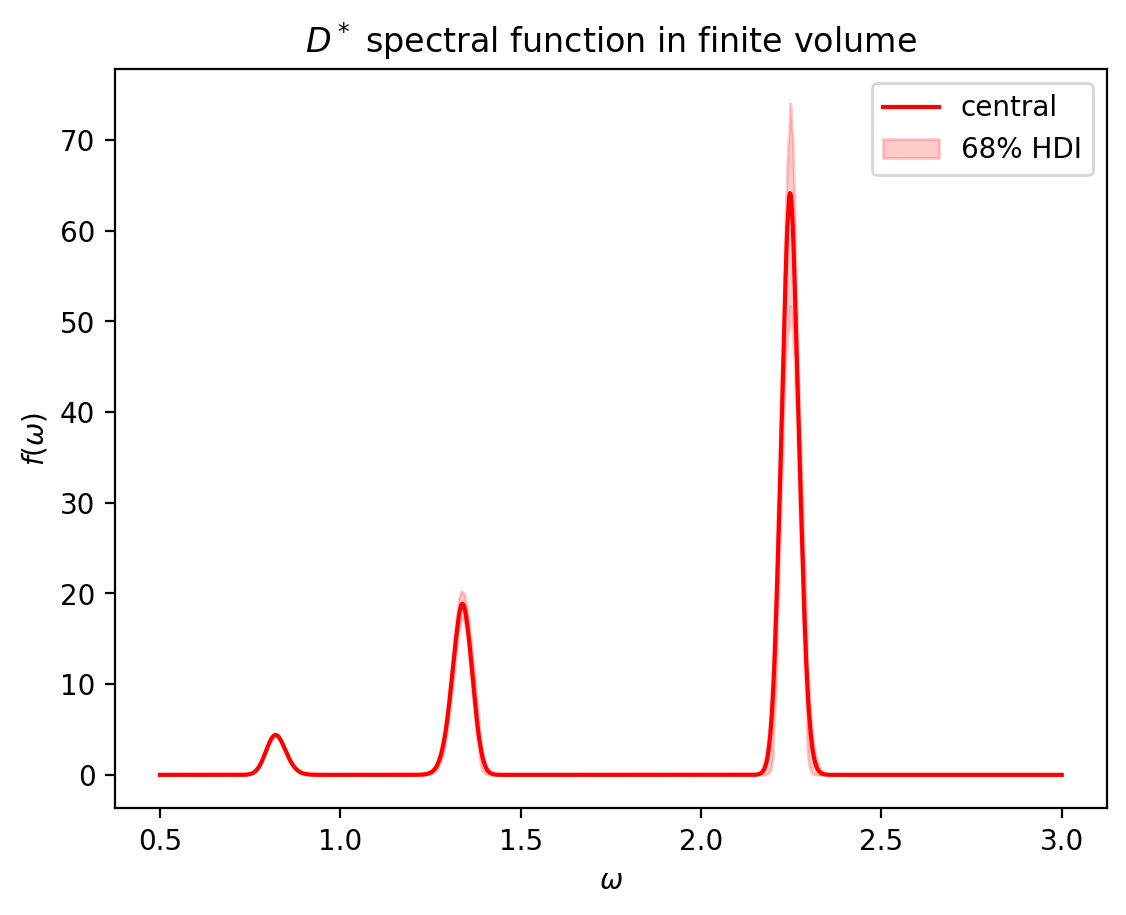}
    \caption{We show the Posterior Predictive Distribution of the spectral function, where three finite-volume states are very clearly visible and can be measured to high-precision by locating the maxima of the peaks for each MCMC sample. We chose the median as central value.}
  \end{subfigure}
  \caption{We choose to apply the GP prior to the latent variable $log(f)$ so that the spectral function $f$ is by construction positive everwhere.
    Using constrains from both positivity and smoothness we obtain very precise results for the spectral function.}
  \label{fig:LatentGP}
\end{figure}

As an alternative to multi-exponential inferences such as Sec.~\ref{sec:multiexp}, one can extract the whole spectral function. A model called BR has already been presented by \cite{Rothkopf}. 
In Fig.~\ref{fig:LatentGP} we propose a more efficient model based on (Hilbert Space) Gaussian Processes. It keeps the positivity constraint of BR while adding some
automatically-adjusted smoothing radius. 
These models are essentially an infinite sum of exponentials, but are more stable than multi-exponential models because no free parameter appear in the exponential: instead the exponentials are applied by a linear operator
\begin{equation}
  {\cal L}\left[f\right](t) = \int K(t,\omega) f(\omega), \qquad K(t,\omega) = \exp\left(-\omega t\right).
\end{equation}

As infinitely many energies are included (up to a large cutoff), such models can typically be fitted on the whole correlator without having to discuss the choice of a fit interval nor
a truncation of the spectrum. Another consequence is that the ground truth is always included in the space covered by the model: this has to be compared to the situation of
multi-exponential $\chi^2$ fits, where the ground truth is almost-surely not included in a finite set of truncated models and which tend to become numerically unstable as we approach the ground truth.

If applied on lattice data in a single finite volume, this spectral function will not have much physical meaning, but it already provides an interesting black box to extract reliably a large number of eigenstates (defined for instance as the modes of the spectral function, sample-per-sample). 

\section{Conclusion}

We presented a well-defined bayesian framework to provide rigorous LQCD results through the whole chain of analysis. 
Its practical applicability has been demonstrated on a large range of models, covering most of the challenges routinely encountered in LQCD.
This brings together a large number of seemingly unrelated issues, which have traditionally been tackled separately by the community: 
resampling and non-gaussianity, $\chi^2$ fits, regularisation of the GLS covariance, auto-correlations, model averaging, inverse problems, \dots

We hope this could constitute a starting point for the community to explore more complicated models and take advantage of these connections.
We invite the reader to have a look at the source code in \cite{zenodo}.
Non-parametric models in particular should be explored further, as the example of the spectral functions shows that a paradigm shift could happen.
This would be consistent with the general trend of machine-learning to move towards universal approximators.

Finally, we focused our discussion on the use of the HMC algorithm, but variational methods would deserve deeper exploration as well, from SVI to neural normalising flows.
Each PPL usually allows several such alternatives to be used interchangeably.

%

\bibliographystyle{apsrev}
\bibliography{proceedingsJFlat23}

\end{document}